\documentclass[useAMS,usenatbib,letterpaper]{mn2e}
\usepackage{psfig}
\newcommand{\beq}{\begin{equation}}
\newcommand{\eeq}{\end{equation}}
\begin{document}
\title[Chemistry and clumpiness in planetary nebulae]{Chemistry and clumpiness in planetary nebulae}
\author[M.P. Redman, S. Viti, P. Cau, D.A. Williams]
{M.P. Redman$^{1}$, S. Viti$^{2}$, P. Cau$^{3,4}$ and D.A. Williams$^{1}$\\
$^{1}$ Department of Physics \& Astronomy, University College London,
Gower Street, London WC1E 6BT, UK\\
$^{2}$ Istituto di Fisica dello Spazio Interplanetario, CNR, Via del Fosso Caveliere 100, 00133 Roma, Italy\\
$^{3}$ Observatoire de Paris-Meudon, 92195 Meudon, France\\
$^{4}$ Department of Meteorology, University of Reading, Earley Gate, P.O. Box 243, Reading RG6 6BB, UK}
\date{\today}
\pubyear{2003} 
\volume{000}
\pagerange{\pageref{firstpage}--\pageref{lastpage}}
\maketitle 
\label{firstpage}

\begin{abstract}
We study the chemistry in the slow wind during the transition from AGB
star to pre-planetary nebula (PPN) and planetary nebula (PN). We show
that there is a very rich chemistry of degradation products created by
photoprocessing, driven by the gradually hardening radiation field of
the central star. Most of these products are, however,
photodissociated during the PPN phase if the wind is smooth. By
contrast, if the wind is clumpy, possibly because of clumpiness in the
AGB atmosphere, then many of these degradation products survive into
the PN phase. Thus, chemistry may be used to infer the existence of
clumpiness in the AGB phase. We identify potential molecular tracers,
and we note that, in the case of clumpiness, large molecules may
survive the transport from the stellar atmosphere to the interstellar
medium. We compare between our model results with observations of
three objects at differing evolutionary stages: CRL618, NGC 7027 and
the Helix nebula (NGC 7293).
\end{abstract}


\section{Introduction}
Nearby planetary nebulae are clearly observed to be clumpy
\citep{odell.et.al02} with perhaps the best known example being the
Helix nebula \citep{meaburn.et.al98b,odell&handron96}. The latter
authors estimate that around half of the mass of the Helix nebula is
contained in clumps. The origin of the clumps remains very unclear and
there have been several models proposed for their
generation. \citet{dyson.et.al89} suggested that the SiO maser spots
observed in the atmospheres of AGB stars could be the first
manifestation of clumps later observed in the planetary nebula. Other
models involve clumps forming much later during the planetary nebula
phases. \citet{williams99} has shown how density fluctuations at an
ionization front boundary can produce knots and tails when the
ionization front switches locally from R-type to D-type (see
e.g. \citealt{shu91} for a detailed discussion of ionization front
types). Instabilities in the shell of ionized gas, with the instability
being either Rayleigh-Taylor \citep{capriotti73} or `thin layer'
\citep{garcia_segura&franco96}, have also been proposed as mechanisms
to form clumps.

Very recently, \citet{huggins&mauron02} have observed the envelopes of
NGC 7027, the proto-typical young planetary nebula, and IRC+10216, the
standard C-rich AGB star, in order to look for knots generated at the
AGB phase. Since none were observed in these two objects, any envelope
knots must be very small or else they would be seen in scattered
light. However, in the innermost regions of NGC 7027, the ionized and
neutral material surrounding the ionized cavity are observed to be
very clumpy and \citet{huggins&mauron02} suggest that the knots are
formed as the complex (and poorly understood) transition from AGB wind
to planetary nebula occurs. The models and explanations for the origin
of the clumps can thus be separated into two types: those in which the
clumps form early, before the PN is fully developed
\citep{dyson.et.al89,huggins&mauron02} and those in which the clumps
form later during the PN stage
\citep{williams99,garcia_segura&franco96,capriotti73}.

The chemical history of the clumps will depend on where and when they
formed since the chemistry is sensitive to the radiation field,
temperature and extinction of the clump gas. Molecular observations
may thus offer the best way to establish the evolutionary history of
the clumps and thus to differentiate between the two types of
model. There has been rapid progress in recent years in characterising
the chemistry of the post-AGB phases with many molecules being
identified in young objects such as NGC 7027
(e.g. \citealt{hasegawa&kwok01} and references therein) and several
species observed in evolved PNe such as the Helix where, for example,
\citet{young.et.al97} observed neutral carbon in the
globules. \citet{bachiller.et.al97} observed molecules in clumps in a
range of planetary nebulae at different evolutionary stages, allowing
trends in the molecular abundances with age to be identified.

In this paper, we trace the chemical evolution of model clumps,
assuming them to have formed early on in the history of the PN and
compare the results from those expected if the clumps formed {\it in
situ}. The aim is to identify observable species that can be used as
discriminants between the two scenarios.

Our model is described in Section 2, and our results are presented,
discussed and compared with observations in Section 3. We conclude in
Section 4 that molecular tracers do exist in PPNe and PNe, and that
some large molecules originating in the stellar atmosphere may survive
the transition into the interstellar medium.

\section{Physical and chemical model}
The difference between a clumpy and non-clumpy model is that clump
interiors will have a higher extinction than their surroundings. In
the harsh environment of a planetary nebula this can help to shield
molecules so that they survive for longer than if they were in the
interclump gas. The chemical difference between clumps that are formed
out of the planetary nebula gas and those that formed earlier is that
in the latter case, complex molecules may be shielded and preserved
long enough to be detectable in clumps. In the former case, all such
molecules will have been destroyed as the extinction of the gas
dropped as the PN evolved.

\citet{howe.et.al92} investigated the gas-phase chemistry in a carbon-rich 
AGB wind during its transition to a planetary nebula. This work was
followed by a paper \citep{howe.et.al94} investigating the formation
of molecules in a dense, neutral globule such as those observed in the
Helix nebula. \citet{howe.et.al94} ran equilibrium chemistry models
for clumps with extinctions between 0 and 2.0, comparable to those of
the Helix knots \citep{meaburn.et.al98b}. They found that molecular
abundances in the globule are enhanced with ${\rm C_2H}$ and ${\rm
CN}$ at abundances that are possibly detectable in carbon-rich
globules.

Here, we extend the work of \citet{howe.et.al94} in that the
time-dependent chemistry is calculated, beginning with the rich
chemistry of a carbon-rich AGB atmosphere. The molecular species are
assumed to be locked into a clump with a density contrast with the
surrounding medium. The chemistry of the clump is then traced as both
it and the surrounding medium expand away from the star. At the final
stage of the calculation, the clump and medium have similar properties
to those of the Helix. As the knots in the Helix have a low
extinction, the chemistry by this stage is tending towards that
expected of a diffuse cloud with this initial composition. However, at
intermediate stages, the rich initial chemistry should give rise to
marked differences compared with that expected from a standard ISM
mixture. Some of these molecules should be present in detectable
quantities and if found, would argue for an early clump formation
model. We follow the physical and chemical evolution of a parcel of
gas (either clump or interclump material) as it moves out from the AGB
atmosphere to the protoplanetary (PPN) and planetary nebula (PN)
phases, ultimately merging with the interstellar medium.

In the AGB atmosphere, the number density and temperature are on the
order of $10^{12}~{\rm cm^{-3}}$ and $10^{3}~{\rm K}$,
respectively. We assume that the clump/interclump density ratio is
about 10 (\citealt{mauron&huggins00} find the multiple shells in the
envelope of IRC+10216 to have density contrasts of $\la 10$), and
that the size of the clump is comparable to a stellar radius (about
$10^{13}~{\rm cm}$). Thus, if the gas:dust ratio is similar to that in
the interstellar medium, the visual extinction associated with a clump
in the atmosphere is very large. The chemistry in the atmosphere is
determined under thermodynamical equilibrium, and the relative
abundances are fixed as the envelope expands and the density falls. At
lower densities, however, thermodynamic equilibrium no longer applies
and chemistry is driven by two-body reactions of ion-molecule and
other types, as in conventional interstellar chemistry.

Since no chemical evolution occurs in the initial expansion, we begin
our computation of the evolution at a later stage (time $t = t_1 =
100~{\rm yr}$) at which the densities are low enough to be dominated
by two-body reactions initiated by cosmic ray ionisation and by
photoprocesses driven by the radiation field of the central star and
by the ambient interstellar medium. 

For simplicity, we use a power-law description of the physical
development. We do not attempt to model the initiation and evolution
of a clump. In a steady outflow at constant velocity, the density
should fall off as $t^{-2}$. However, the density fall off in a clump
may be slower. A $t^{-3/2}$ dependence gives a reasonable fit to
typical densities through the PPN and PN stages (as observed by,
e.g. \citealt{martin-pintando.et.al95}; \citealt{meaburn.et.al98b}).
Therefore we adopt
\beq
n(t)=10^7~\left( \frac{t}{t_1} \right)^{-3/2}~{\rm cm^{-3}},
\eeq
for times $t > t_1$, where $t_1 = 100~{\rm yr}$. On the assumption of
constant clump mass throughout the expansion, this gives
\beq
r(t)=10^{14}\left( \frac{t}{t_1} \right)^{1/2}~{\rm cm}.
\eeq
The visual extinction associated with a clump must therefore vary as 
$t^{-1}$, and we write 
\beq
A_{V}=10^2\left( \frac{t}{t_1} \right)^{-1}.
\eeq
The distance of the clump from the star is $d$, which in a steady flow must 
increase linearly with time:
\beq
d(t) = 10^{16}\left( \frac{t}{t_1} \right)^{}~{\rm cm}.
\eeq
We adopt the following dependence of temperature on time, as it gives
values for the PN clump temperatures comparable to the typical measured values
\citep{martin-pintando.et.al95,meaburn.et.al98b}:
\beq
T(t) = 316\left( \frac{t}{t_1} \right)^{-1/4}~{\rm K}.
\eeq
The radiation field experienced by the parcel of gas is initially
dominated by the contribution from the central star, and later by that
of the interstellar radiation field. We adopt the following
expression, similar to that used by \citet{howe.et.al94}, for the
radiation field intensity $\chi$,
\begin{eqnarray}
\chi & = & 0;~D<10^{16}~{\rm cm}, \\
& = & 100\chi_{\rm ism} \left( \frac{t}{t_1} \right)^{-2} +
\chi_{\rm ism};~D<10^{16}~{\rm cm},
\end{eqnarray}
where we assume that $\chi_{\rm ism}$ gives photorates as in the UMIST
rate file (\citealt{millar.et.al97}, and discussed further
below). There is some uncertainty in this, since the radiation field
may be harder than that of the interstellar medium. We use the
canonical cosmic ray ionisation rate of $\chi_{\rm ism}=1.3 \times
10^{-17}~{\rm s^{-1}}$.

The evolution of the interclump medium can be followed in a very
similar way to that of a clump. In this single-point calculation, the
initial composition in the AGB atmosphere is the same as for a clump
but the parcel of gas is assumed to have an initial density lower by a
factor of 10 compared with the clump. The density falls off with the
square of the distance from the star, as appropriate for a steady
spherically symmetric flow.
\beq
n(t)=10^6~\left( \frac{t}{t_1} \right)^{-2}~{\rm cm^{-3}}.
\eeq
A parcel of gas will then experience an extinction that varies as
\beq
A_{V}=10^2\left( \frac{t}{t_1} \right)^{-4/3}.
\eeq
Since the clump and interclump medim will initally have the the same
temperature, it is assumed for simplicity that the interclump medium
temperature is the same as that of the clump, and therefore varies as
given by Equation~5. This is unlikely to be true in the later stages
of the PN evolution when the interclump medium will be hotter than the
clumps but this will only accelerate the destruction of those
molecules that survived the drop in extinction. In the period of
interest, the range of temperatures involved does not influence the
chemistry very strongly.

The chemical model used is similar to the one employed in
\citet{viti&williams99}. However, we have made considerable changes
to the chemistry.  To all the basic species (O, C, CO etc.)  we have
added all the species (mainly hydrocarbons) that are thought to be
abundant at the initial stage of the PPN; this chemistry also includes
species that have an unpaired electron located on an internal or
terminal carbon atom. For most of these additions, the UMIST rate file
for gas phase reactions, involving several hundred species
\citep{millar.et.al97}, and which is normally used routinely in our models, did
not have a sufficient network of reactions. Therefore the formation
and destruction routes were taken from
\citet{frenklach&feigelson89}. The reaction rates were fitted by using
different formulae from the UMIST ones. To account for that, we have
modified our code to include new rate coefficients calculations taken
from \citet{cau01}. For some of those reactions, no rate coefficients
are available and
\citet{frenklach&feigelson89} determine them from the rate
coefficients of their inverse reactions: we have employed the same
method here.  We have not included three-body reactions.  We have also
excluded reactions which use an `unspecified' species as a collisional
body causing de-excitation or dissociation.  These may be of some
relevance and should be included in further studies.  Note that
\citet{frenklach&feigelson89} also extrapolated their rate coefficients to
temperatures very different from those at which they were measured.

\section{Results and discussion}
\subsection{Predictions}
We present in Figures~1-4 results of the calculations of the evolution
of molecular fractional abundances in clump and interclump gas, for
several illustrative species. In Table~1 we give the
computed fractional abundances of potentially observable species in
the clump, and also the clump:interclump ratio of abundances, for
several epochs between about 2 and 10 thousand years, i.e. the period
of transition from PPN to PN phases.

The figures show generally the same qualitative behaviour. A comparison of 
clump and interclump abundances shows that molecules are photodissociated 
earlier in the interclump than in the clump gas, as expected. Thus 
molecules survive longer in the clumps, and so the ratio of clump to 
interclump abundances varies strongly. For example, the CN abundance ratio 
varies from 0.26 at 2550 yr to 270 at 10050 yr, a factor of 1000. 

\begin{table*}
\begin{tabular}{lllllll}
Species & $X(\rm clump)$ & $\frac{X(\rm clump)}{X(\rm interclump)}$ & $X(\rm clump)$ & $\frac{X(\rm clump)}{X(\rm interclump)}$ & $X(\rm clump)$ & $\frac{X(\rm clump)}{X(\rm interclump)}$ \\ & $t=2550~{\rm yr}$ & $t=2550~{\rm yr}$ & $t=6300~{\rm yr}$ & $t=6300~{\rm yr}$ & $t=10050~{\rm yr}$ & $t=10050~{\rm yr}$\\
\hline
C             &$3.4\times 10^{-6}$  & $1.4\times 10^{-2}$  &$ 3.4\times 10^{-4}$ & 2.4          &$ 6.0\times 10^{-4}$ & 24 \\
${\rm C^+} $  &$ 7.0\times 10^{-11}$  & $6.0\times 10^{-6}$  &$ 1.2\times 10^{-5}$ & $1.1\times 10^{-2}$  &$ 2.7\times 10^{-4}$ & 0.22 \\
CH            &$2.0\times 10^{-8}$  & $8.9 \times 10^{-2}$ &$ 2.6\times 10^{-9}$ & 0.65        &$ 4.4\times 10^{-9}$ & 7.0 \\
${\rm CH^+}$  &$1.8\times 10^{-15}$  & $1.5 \times 10^{-4}$ &$ 5.8\times 10^{-14}$ & $5.1\times 10^{-3}$ & $ 1.8\times 10^{-12}$ & 0.14 \\
CO            &$9.1\times 10^{-4}$  & 1.2               &$ 6.9\times 10^{-4}$ & 18              &$ 1.8\times 10^{-4}$ & 580 \\
CN            &$5.1\times 10^{-6}$  & 0.26                 &$ 1.1\times 10^{-7}$ & 55           &$ 9.4\times 10^{-10}$ & 270\\
CS            &$1.1\times 10^{-5}$  & 3.8                  &$2.0\times 10^{-6}$ & 1300          &$ 1.5\times 10^{-8}$ & $3.3\times 10^{4}$\\
SiC           &$7.0\times 10^{-7}$  & 0.24                 &$1.4 \times 10^{-8}$ & 260          &$ 4.6\times 10^{-11}$ & 2.4\\
HCN           &$4.0\times 10^{-5}$  & 22                   &$ 1.0\times 10^{-10}$ & 190         &$1.3\times 10^{-12}$ & 21\\
HNC           &$1.4\times 10^{-8}$  & 22                   &$ 2.7\times 10^{-12}$ & 5.3                 &$ 2.2\times 10^{-12}$ & 70\\
HCSi          &$5.3\times 10^{-8}$  & $6.3\times 10^{-2}$  &$1.8\times 10^{-8}$ & 2200          &$ 6.2\times 10^{-11}$ & 190\\      
${\rm HCO^+}$ & $1.1\times 10^{-10}$  & 0.27 & $4.4\times 10^{-12}$ & 0.88 & $ 6.1\times 10^{-12}$ & 7.1\\
${\rm SiCH_2}$&$1.6\times 10^{-8}$  & 7.9                  &$3.0\times 10^{-12}$ & $6.5\times 10^5$    &$ 1.2\times 10^{-16}$ & 500\\
${\rm C_2}$   &$1.0\times 10^{-6}$  & $2.5\times 10^{-2}$  &$ 2.6\times 10^{-7}$ & 1.7                 &$ 1.0\times 10^{-8}$ & 45\\
${\rm C_2H}$  &$9.7\times 10^{-6}$  & 0.47                 &$ 3.8\times 10^{-8}$ & 18           &$ 6.7\times 10^{-10}$ & 79\\
${\rm C_2H_2}$&$2.1\times 10^{-7}$  & 7.0                  &$ 3.2\times 10^{-10}$ & 610                 &$2.9 \times 10^{-12}$ & 86\\
${\rm SiC_2}$ &$6.4\times 10^{-6}$  & 0.30                 &$ 1.2\times 10^{-7}$ & 1200                &$6.9 \times 10^{-11}$ & 2.0\\
${\rm SiC_2H}$&$ 2.8\times 10^{-8}$  & 4.0                  &$ 1.1\times 10^{-10}$ & $5.4\times 10^5$    &$ 1.6\times 10^{-15}$ & 1600 \\
${\rm SiC_2H_2}$&$2.4\times 10^{-8}$& 4.8                  &$ 9.4\times 10^{-11}$ & $7.6\times 10^6$    &$ 1.4\times 10^{-16}$ & 470\\
${\rm C_3}$    &$1.2\times 10^{-7}$ & 0.28                 &$ 9.7\times 10^{-8}$ & 430                 &$ 5.9\times 10^{-10}$ & 1400\\
${\rm C_3H}$  &$7.1\times 10^{-7}$  & 3.0                  &$ 1.0\times 10^{-9}$ & 460                 &$1.3 \times 10^{-12}$ & 62\\
${\rm HC_3N}$ &$3.7\times 10^{-6}$  & 4000                 &$ 1.4\times 10^{-12}$ & $3.9\times 10^4$    &$ 1.1\times 10^{-15}$ & $1.7\times 10^5$\\
${\rm SiC_3}$ &$1.2\times 10^{-7}$  & 0.85                 &$ 1.4\times 10^{-8}$ & 110                 &$ 4.3\times 10^{-13}$ & $7.0\times 10^{-3}$\\
${\rm C_4}$   &$1.7\times 10^{-9}$  & $2.4\times 10^{-3}$  &$ 1.3\times 10^{-7}$ & 300                 &$6.0 \times 10^{-12}$ & 3.4\\
${\rm C_4H}$  &$6.0\times 10^{-9}$  & $9.0\times 10^{-3}$  &$ 1.6\times 10^{-8}$ & 450                 &$8.9 \times 10^{-12}$ & 92\\
${\rm SiC_4}$ &$1.0\times 10^{-7}$  & 3.0                  &$4.2\times 10^{-11}$ & $5.1\times 10^{-3}$ &$5.1\times 10^{-18}$ & $1.2\times 10^{-9}$ \\
${\rm C_4H_2}$&$2.2\times 10^{-9}$  & $4.7\times 10^{-3}$  &$1.5\times 10^{-10}$ & 270          &$1.2\times 10^{-12}$ & 22\\
${\rm C_4H_3}$&$ 6.4\times 10^{-6}$  & 0.52                 &$ 3.1\times 10^{-9}$ & 3900                &$ 4.1\times 10^{-12}$ & 90 \\
${\rm C_4H_4}$&$ 2.2\times 10^{-8}$  & $6.1\times 10^{-2}$  &$ 3.0\times 10^{-6}$ & 0.61                &$ 1.5\times 10^{-5}$ & 1.8\\
${\rm C_4H_5}$&$6.6\times 10^{-5}$  & 7.0                  &$ 6.3\times 10^{-5}$ & 13                  &$ 5.1\times 10^{-5}$ & 33 \\
${\rm C_5}$   &$4.6\times 10^{-10}$  & $2.6\times 10^{-3}$  &$ 1.1\times 10^{-7}$ & 160                 &$6.7 \times 10^{-12}$ & 1.7\\
${\rm C_5H}$  &$1.5\times 10^{-8}$  & $1.4\times 10^{-2}$  &$ 7.5\times 10^{-9}$ & 2400                &$5.9 \times 10^{-13}$ & 39\\
${\rm C_5N}$  &$4.0\times 10^{-11}$  & $2.5\times 10^{-3}$  &$ 1.4\times 10^{-8}$ & 7900                &$5.7 \times 10^{-14}$ & 1600\\
${\rm HC_5N}$ &$ 9.4\times 10^{-10}$  & $5.3\times 10^{-2}$  &$ 3.8\times 10^{-13}$ & $2.4\times 10^4$    &$ 1.6\times 10^{-17}$ & 3600 \\
${\rm C_6}$   &$4.1\times 10^{-10}$  & $5.5\times 10^{-3}$  &$ 2.4\times 10^{-8}$ & 130                 &$ 1.9\times 10^{-13}$ & 0.13\\
${\rm C_6H}$  &$1.0\times 10^{-7}$  & 1.1                  &$ 2.7\times 10^{-8}$ & 34                  &$ 3.0\times 10^{-14}$ & $4.9\times 10^{-3}$ \\
${\rm C_6H_2}$&$1.8\times 10^{-8}$  & 40                &$ 7.0\times 10^{-17}$ & 87             &$ 1.1\times 10^{-18}$ & 0.22\\
Benzene       &$2.6\times 10^{-8}$  & 90                  &$ 2.6\times 10^{-8}$ & 90                 &$2.6 \times 10^{-8}$ & 90\\
Benzyne       &$1.9\times 10^{-8}$  & 7.1                   &$ 1.9\times 10^{-8}$ & 7.1                  &$1.9 \times 10^{-8}$ & 7.1\\
${\rm C_7}$   &$1.7\times 10^{-10}$  & $1.1\times 10^{-2}$  &$6.0\times 10^{-9}$ & 630                 &$ 1.1\times 10^{-14}$ & 0.49\\
${\rm C_7H}$  &$5.0\times 10^{-8}$  & 1.6                  &$7.7\times 10^{-12}$ & $1.2\times 10^4$    &$2.9\times 10^{-18}$ & 6.5 \\
${\rm HC_7N}$ &$ 2.8\times 10^{-8}$  & 6.1                  &$ 1.5\times 10^{-15}$ & $2.4\times 10^6$    &$ 3.4\times 10^{-23}$ & 33\\
O             &$2.0\times 10^{-6}$  & $1.2\times 10^{-2}$  &$2.2\times 10^{-4}$ & 0.25          &$ 7.3\times 10^{-4}$ & 0.80\\
Si            &$2.7\times 10^{-5}$  & 12                   &$5.0 \times 10^{-7}$ & 5.3                 &$ 2.3\times 10^{-7}$ & 4.2\\
SiO           &$3.4\times 10^{-8}$  & 0.14                 &$ 1.9\times 10^{-6}$ & $1.2\times 10^5$    &$ 1.1\times 10^{-10}$ & 46\\
S             &$1.4\times 10^{-6}$  & 0.12                 &$ 2.4\times 10^{-6}$ & 26                  &$ 1.6 \times 10^{-7}$ & 2.3\\
SiS           &$9.6\times 10^{-6}$  & 2.9                  &$ 1.3\times 10^{-14}$ & $3.8\times 10^4$    &$ 1.2\times 10^{-17}$ & 200 \\
${\rm S^+} $  &$2.5\times 10^{-10}$  & $7.5\times 10^{-5}$  &$ 1.7\times 10^{-5}$ & 0.80                &$ 2.2\times 10^{-5}$ & 0.99\\
\hline
\end{tabular}
\label{compare}
\caption{Comparison of fractional abundances of potentially observable species at
different times. The density, temperature and extinction at these
times can be readily calculated from the equations in Section 2}
\end{table*}

\begin{figure}
\psfig{file=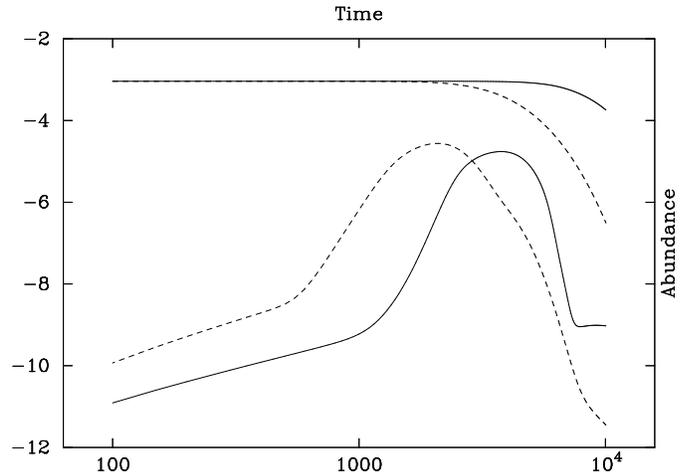,width=250pt,bbllx=17pt,bblly=0pt,bburx=739pt,bbury=510pt}
\caption{Clump (solid lines) and interclump (dashed lines) abundances for CO
(uppermost solid line) and CN. The time is measured in years from the
end of the AGB phase. The fractional abundance is with respect to
the total number of hydrogen atoms. The object becomes a PPN after a few
hundred years and by about 1000 years the object is a PN}
\label{species6}
\end{figure}
\begin{figure}
\psfig{file=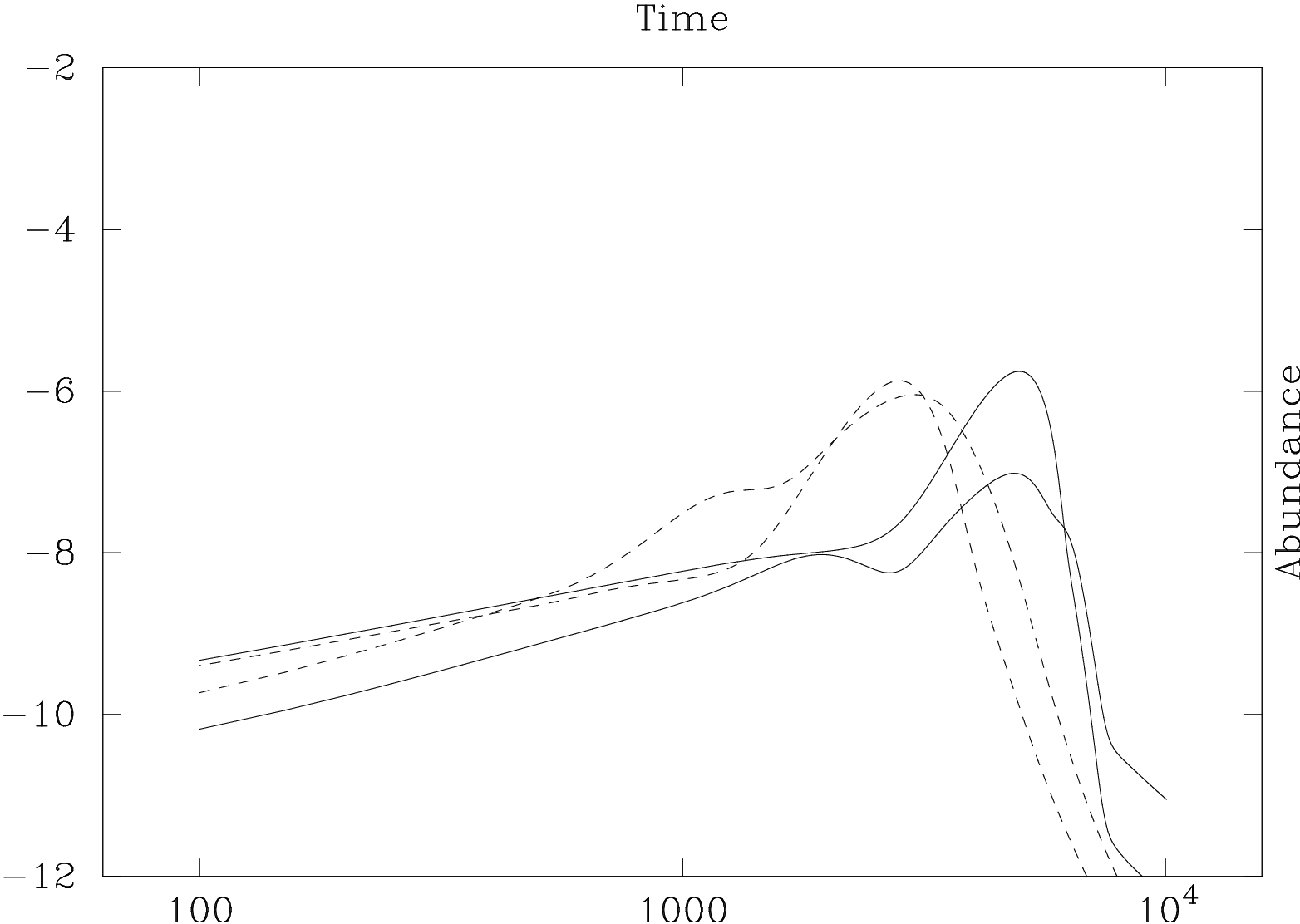,width=250pt,bbllx=17pt,bblly=0pt,bburx=739pt,bbury=510pt}
\caption{Clump (solid lines) and interclump (dashed lines) abundances for ${\rm
C_5H}$ (uppermost solid line) and ${\rm C_4H}$. The time is measured
in years and the fractional abundance is with respect to the total
number of hydrogen atoms.}
\label{species8}
\end{figure}
\begin{figure}
\psfig{file=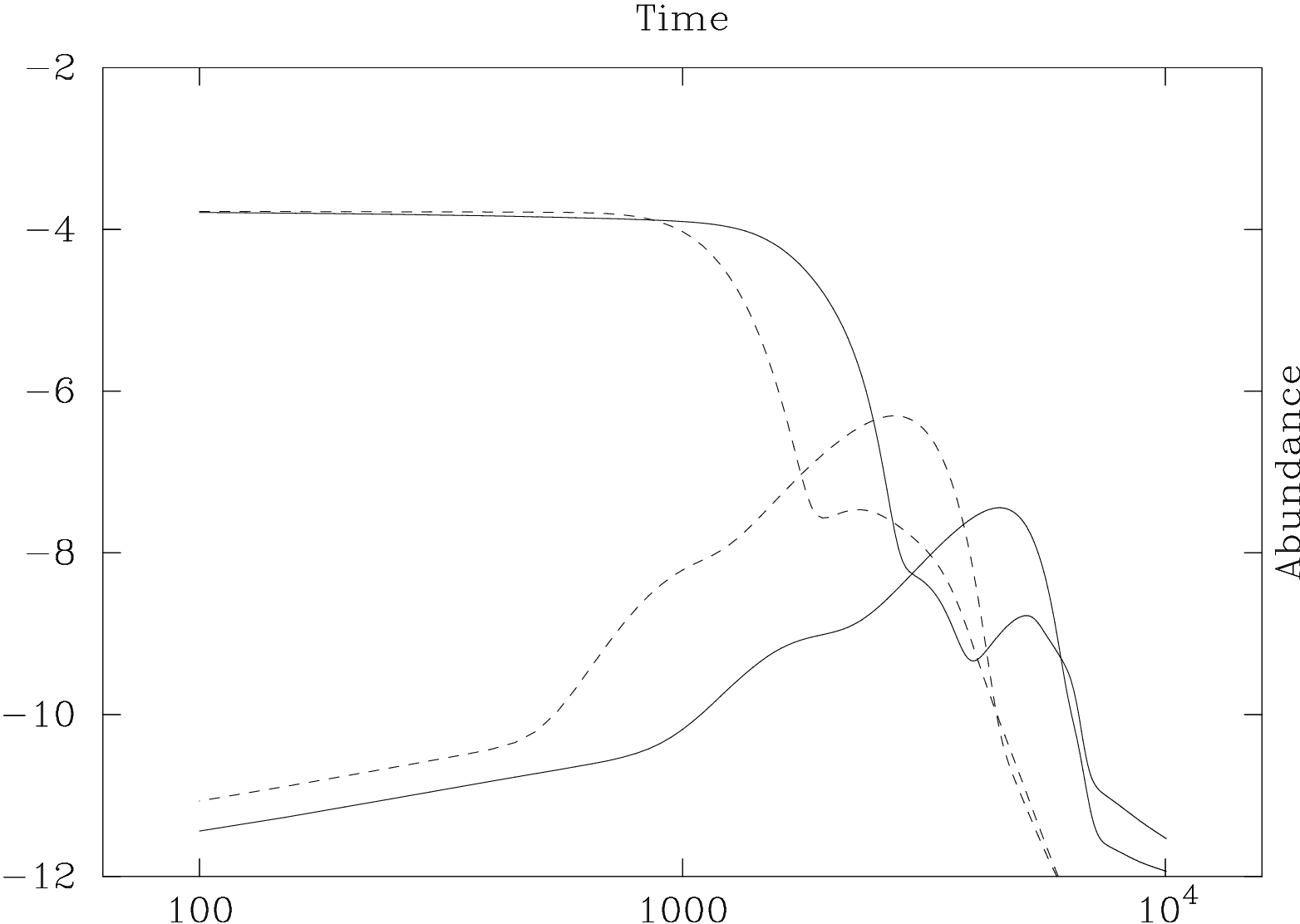,width=250pt,bbllx=17pt,bblly=0pt,bburx=739pt,bbury=510pt}
\caption{Clump (solid lines) and interclump (dashed lines) abundances for ${\rm
C_2H_2}$ (uppermost solid line) and ${\rm C_4H_2}$. The time is
measured in years and the fractional abundance is with respect to the
total number of hydrogen atoms.}
\label{species1}
\end{figure}
\begin{figure}
\psfig{file=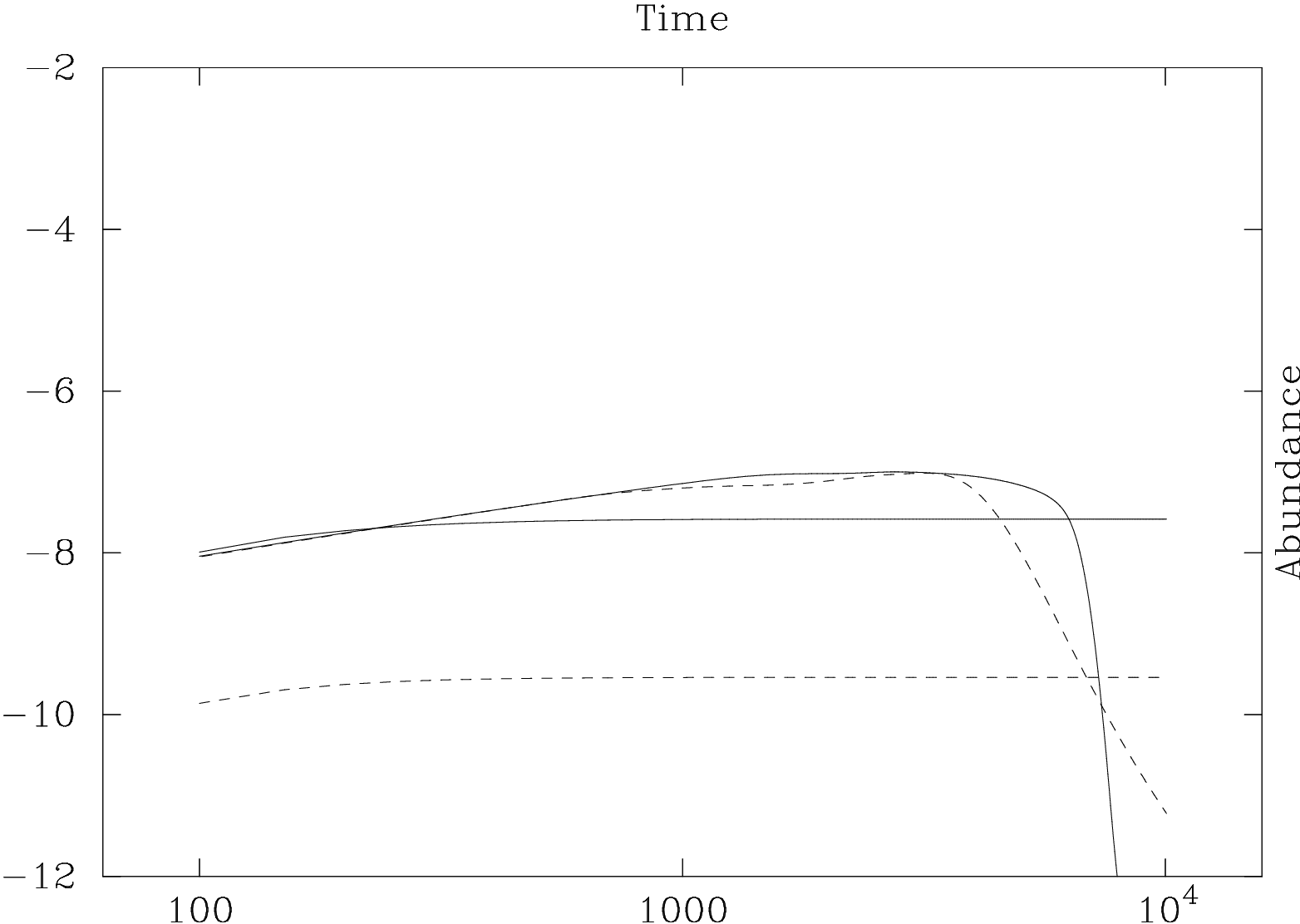,width=250pt,bbllx=17pt,bblly=0pt,bburx=739pt,bbury=510pt}
\caption{Clump (solid lines) and interclump (dashed lines) abundances for ${\rm
C_6H}$ (uppermost solid line) and benzene. The time is measured in
years and the fractional abundance is with respect to the total number
of hydrogen atoms.}
\label{species4}
\end{figure}

Figure 1 shows the clump and interclump evolution of CO and CN
fractional abundances. These represent molecules that have high and
low initial abundances, respectively, i.e. in the stellar atmosphere
of the AGB star.  The abundance of CO remains high in both clump and
interclump gas until photodissociation begins to play a role as the
gas density and extinction decline during the expansion, with CO in
the higher density clump gas surviving until later times. On the other
hand, CN is a molecule formed as a product of photodissociation of
larger species, here called a {\it degradation product}. Since
photodissociation is faster in the interclump gas, the production of
CN is also faster there initially, so CN rises faster in the
interclump gas than in the clump, but also declines earlier.

Similar effects are observed for other species. The simple carbon
chains ${\rm C_4H}$ and ${\rm C_5H}$ are both degradation products of
larger species, and the time evolution illustrated in Figure 2 shows
that their abundances peak in the interclump gas earlier (by several
thousand years) than in the clump gas. Figure 3 shows the time
evolution of ${\rm C_2H_2}$ (which is abundant in the stellar
atmosphere) and ${\rm C_4H_2}$, a degradation product of larger
species. Their behaviours are as for CO and CN. Both molecules
ultimately decline as the gas enters the interstellar medium.

Molecules with rather slower photodissociation rates survive
longer. In Figure 4, benzene has a timescale for survival longer than
the ten thousand year period considered here. This raises the
possibility that some large and relatively photo-resistant species may
survive the transition into the interstellar medium and may establish
a population there (Williams 2003). Note however that our benzene
chemistry may be incomplete; in the chemistry adopted here its
formation and destruction are mainly dominated by the neutral-neutral
reaction of two ${\rm C_3H_3}$ molecules and its reverse.

Table~1 illustrates the effects of the different
photodissociation timescales in the expanding gas, and gives snapshot
values of the clump:interclump abundance ratio at 2550, 6300, and
10050~yr, for species that are potentially observable. The earliest
epoch is approximately when the interclump abundances are at a
maximum, while the values in the clump of species formed by
degradation are low. For such species, then the ratio is low.

At epoch ${\rm 6300~yr}$, many of the degradation products have
declined in abundance in the interclump gas, while their abundances in
the clump gas are growing, so the ratio is generally large. At the
latest epoch shown in Table~1, then most of the degradation products
have declined in abundance in the clump gas, and have declined even
further in the interclump gas, so the ratio is in many cases again
large. However, by this stage the abundances in the clump gas may have
declined so far as to make many of the species undetectable. If we
take a fractional abundance of $10^{-8}$ to represent a characteristic
detection limit at ${\rm 10000~yr}$, where the clump extinction is
$\sim 1$ magnitude, then about 70\% of the species listed are possibly
not detectable at this epoch, whereas only about 30 percent are not
detectable at ${\rm 6500~yr}$. Thus, there is a very significant
change in the detectable chemistry of the clump gas the PN ages.

The column density of any species is made up of contributions from
both clump and interclump gas. At the earliest PPN epoch illustrated,
Table~1 shows that chemistry is largely determined by the
interclump gas, while at later epochs it is determined in the
clumps. Table~2 lists molecules that are likely to be
the most important tracers of either clump or interclump gas, at the
three epochs chosen.

\begin{table}
\begin{tabular}{lll}
& Clump & Interclump \\
\hline
$X>10^{-6}$ at 2550~yr  & CO, CS, HCN, ${\rm C_3H}$, &  CO, CN, SiC, ${\rm C_2}$, \\
 & ${\rm HC_3N}$, ${\rm C_4H_5}$, SiS  &  ${\rm C_2H}$, ${\rm SiC_2}$, ${\rm C_4H_3}$  \\
& & \\
$X>10^{-7}$ at 6300~yr  & CO, CS, HCN, ${\rm C_3H}$, & ${\rm C_4H_4}$ \\
 & ${\rm HC_3N}$, ${\rm C_4H_5}$, SiS   &   \\
& & \\
$X>10^{-8}$ at 10050~yr & CS, ${\rm C_2}$, ${\rm C_4H_4}$, ${\rm C_4H_5}$ & none \\
\hline
\end{tabular}
\label{strongest}
\caption{
List of important potential tracers of the clump and interclump
gas at the three epochs}
\end{table}

\subsection{Comparison with observations}
Here the results will be compared very generally with observations of
three objects: CRL618, NGC7027 and the Helix nebula. These objects
span progressively advanced evolutionary stages with CRL618 being a
PPN, NGC 7027 a young PN and the Helix an evolved PN. As discussed
immediately below, comparison with CRL618 is made difficult by the
complex geometry of the source. The comparisons with NGC 7027 and the
Helix are more straight-forward.

Carbon-rich protoplanetary nebulae such as CRL618 and AFGL2688 have
been the subject of several molecular studies in recent years.  In
particular, CRL618 has been used as a typical example of an asymptotic
giant branch (AGB) star in the transition evolving toward the
planetary nebula stage (\citealt{herpin&chernicharo00}; Cernicharo et
al. 2001a,b; Woods et al. 2002).  Here we attempt a qualitative
comparison of our models with some observations and theoretical work
on this object. \citet{herpin&chernicharo00} show that to reproduce
the observed emission of CO, HCN, HNC, H$_2$O, OH and O, a 3-component
geometry is needed (see their figure 3), with a central torus at high
density ($\sim 5 \times 10^7~{\rm cm}^{-3}$) and a gradient of
temperatures ranging from 250 to 1000 K; an extended AGB remnant
envelope with a density of $5\times 10^5~{\rm cm^{-3}}$ and
temperature of 100K; and finally the lobes, where high velocity gas is
emitted from, with a density of $10^7~{\rm cm^{-3}}$ and a
two-component temperature of 200 and 1000 K.  In follow up papers,
Cernicharo et al.\@~(2001a,b) observed methylpolyynes, small
hydrocarbons and benzene along the same line of sight and conclude
that the emission of these species must come from the central torus
region.

The scenario we model is different from the 3-component geometry but
we can attempt a comparison with the species emitted from the AGB
remnant component where the temperature is close to the one derived in
our model (see Table 1), although the density is much higher. We find
that our best match with the PPN CRL618 occurs at early times $\sim $
few hundred years. At this time the density is relatively high: our CO
column density is $6\times 10^{18}~{\rm cm^{-2}}$, while Herpin
\& Cernicharo (2000) derive a value of $7\times 10^{18}~{\rm cm^{-2}}$
and our HCN is $2.5\times 10^{17}~{\rm cm^{-2}}$, only slightly over a
factor of 4 less than the observed one; our HNC is, however, over 3
orders of magnitude underabundant with respect to observations. Given
the differences in the physical parameters adopted, any further
comparison would be of limited value.

It should also be noted that following the ISO discovery of benzene in
CRL618 (Cernicharo et al.\@~2001b), Woods et al. (2002) presented a
chemical model of this proto-planetary nebula. They show that the
physical conditions of CRL618 are such to encourage an efficient
production of benzene. However, similarly to Cernicharo et
al.\@~(2001b) results, they model the dense inner torus and therefore
we are unable to make a direct comparison with their chemical model.

The young PN NGC 7027 is molecule rich with a large number of species
identified within it. \citet{hasegawa&kwok01} have observed several
species and collated the results of previous work. In their table 3,
they list measured fractional abundances. If the emission that
\citet{hasegawa&kwok01} detected is regarded as originating from a
large collection of clumps then their results can be directly compared
with ours in Table 1. Adopting an age of around 1000~yr for the nebula
and comparing with our results gives the following: CN is predicted to
have a fractional abundance of $9\times 10^{-10}$ and
\citet{hasegawa&kwok01} measure $1.7\times 10^{-8}$; HCN is $6\times
10^{-5}$ (predicted) verses $>1.2\times 10^{-9}$ (observed); and ${\rm
C_2H}$ is $6\times 10^{-7}$ vs $1.1\times 10^{-8}$. Given the
simplicity of the model and the inexactness of the comparison to be
within around one order of magnitude of the observational results for
several species is very encouraging (the exact age of the nebula or
the filling factor of the clumps can easily yield uncertainties of an
order of magnitude).

To compare the results for the later stages with those of the Helix,
we can use the last columns of Table 1 with the results of
\citet{bachiller.et.al97}. The model results for this evolved PN do
not agree nearly so well with the observations as for NGC7027 and
CRL618. Some results are not inconsistent with the observations: the
model predictions do not exceed the upper limits established for SiO
and CS, being an order of magnitude less; the predicted ${\rm HC_3N}$
abundance is several orders of magnitude less than the upper limit set
by \citet{bachiller.et.al97}. However the models severely underpredict
the abundances of HCN, HNC, CN and ${\rm HCO^{+}}$. This indicates
that these molecules whose earlier clump abundances were comparable to
the observed levels survive for longer than our simple model predicts.

\section{Conclusions}
The chemistry of the gas in the slow wind ejected from a star in the
AGB phase is strongly affected by photoprocesses driven by the
hardening radiation field of the central star, as the system is
transformed into a PPN and then to a PN. If the slow wind density
distribution is smoothly monotonic, then the degradation product
molecules in the gas are confined to the PPN stage. However, if the
slow wind is clumpy, perhaps because of clumpiness in the precursor
AGB atmosphere, then molecules in clumps are protected against
photodissociation until significantly later times, and may be present
in the PN phase of evolution. We have presented lists of molecules
that may represent tracers of such clumpiness. Benzene survives the
transition into the PN phase, suggesting that some long-lived
molecules may last long enough to be injected into the general
interstellar medium. A comparison with observations suggests that this
model can, at the PPN and young PN stage, correctly predict the
abundance of several molecular species to within an order of
magnitude.  At later stages, the model severely underestimates the
abundance of some short-chain molecules. Thus the shielding of
molecules in clumps may be even more effective than as described in
this simple model. 

The original aim of this work was to chemically identify when clumps
form in PNe. The comparison of the results in the early stages with
the data from CRL618 and NGC 7027 supports the hypothesis that the
clumps form at an early stage; if they formed later out of low
extinction gas, the abundances of the molecules would be significantly
depleted compared with the values observed. As noted above, the
evidence of the continuing presence of molecules in an evolved PN like
the Helix indicates that the protective ability of the clumps persists
until late stages. Future work in this field should include a more
realistic physical model for the different post-AGB phases to account
for the survival of molecules in the clumps. It should also be
possible to model individual sources and thus carry out a much more
detailed comparison with observations both from existing facilities
and also from forthcoming facilities such as ALMA.

\section*{Acknowledgements}
We thank the referee, Pierre Cox, for a very constructive report that
led to an improved paper. We also thank John Dyson and Jonathan
Rawlings for useful discussions. MPR acknowledges the support of
PPARC. SV thanks the Italian Space Agency (ASI) and DAW thanks the
Leverhulme Trust for financial support. PC was a Fellow supported by
the European Commission TMR programme in Astrophysical Chemistry while
this work was carried out.

\label{lastpage}
\end{document}